\documentclass[11pt]{article}

\include{setup}

\begin{document}

\thispagestyle{empty}
\def\thefootnote{\fnsymbol{footnote}}
\setcounter{footnote}{1}
\null
\vskip 0cm
\vfill
\begin{center}
 {\Large \boldmath{\bf Next-to-leading order QCD corrections to Higgs production at a future lepton-proton collider}
\par} \vskip 2.5em
{\large
{\sc B.\ J\"ager$^1$
}\\[2ex]
{\normalsize \it 
$^1$ \it Institut f\"ur Theoretische Physik und Astrophysik, Universit\"at W\"urzburg, 97074 W\"urzburg, Germany
}\\[1ex]
}
\par \vskip 5em
\end{center}\par
\vskip .0cm \vfill {\bf Abstract:} \par
Crucial information on the coupling of the Higgs boson to bottom quarks is expected from Higgs production in association with a forward tagging jet at a future high-energy lepton-proton collider. In order to control the theoretical uncertainties of the signal process, the impact of radiative corrections has to be quantified. We present the full next-to-leading order QCD corrections to $e^-p\to e^-jH$ and  $e^-p\to \nu_ejH$ in the form of a flexible Monte-Carlo program allowing for the calculation of cross sections and kinematic distributions within experimentally feasible selection cuts. 
QCD corrections are found to be very small for cross sections, while the shape distortion of distributions can be as large as 20\%. Residual scale uncertainties at next-to-leading order are at the permille level. 

\par
\null
\setcounter{page}{0}
\clearpage
\def\thefootnote{\arabic{footnote}}
\setcounter{footnote}{0}

\section{Introduction}
With the start-up of the CERN Large Hadron Collider (LHC), promising prospects have arisen for pinning down the mechanism responsible for electroweak symmetry breaking. In order to verify the scenario realized by the Standard Model (SM), a neutral, CP-even Higgs boson has to be discovered and its properties have to be investigated carefully. This implies measurement of its couplings to gauge bosons and fermions. At the LHC, the determination of the Higgs couplings to the top quark, tau lepton, and the weak gauge bosons should be possible, once a sufficient amount of data has been collected \cite{Duhrssen:2004cv, Zeppenfeld:2000td,Rainwater:1998kj}. 
If the Higgs boson turns out to be relatively light, its dominant decay mode is to bottom quarks. Extracting the bottom Yukawa coupling is of particular importance, as this quantity will help to distinguish the SM from various of its extensions such as supersymmetric scenarios. 

However, probing the $Hb\bar{b}$ coupling at the LHC is a challenging enterprise due to large QCD backgrounds in the dominant production modes $gg\to H$, associated $ZH$ and $WH$ production, and $t\bar tH$ production with subsequent decay $H\to b\bar b$ \cite{Duhrssen:2004cv}. Higgs production via weak-boson fusion (WBF) with decay into bottom quarks, on the other hand, provides challenges for the {\tt ATLAS} and {\tt CMS} triggers~\cite{Mangano:2002wn}.  
Therefore, additional search channels, such as WBF production of $\gamma H$~\cite{Gabrielli:2007wf} and $WH$~\cite{Rainwater:2000fm},  with $H\to b\bar b$ have been considered. Alternatively, one could make use of the substructure of so-called ``fat'' jets resulting from the decay products of heavy particles in different production modes~\cite{Butterworth:2008iy,atlas-2009-088,Plehn:2009rk}. 
Despite these recent developments, the observation of the $H\to b\bar b$ decay mode remains challenging at the LHC.

New opportunities could be provided by a future high-energy lepton-hadron collider such as the CERN Large Hadron electron Collider (LHeC), which is currently under scrutiny~\cite{Dainton:2006wd}. For now, various scenarios are being discussed, which all make use of the LHC proton beam with an energy of $E_p= 7$~TeV and an electron beam in the range of $E_e = 50 - 200$~GeV, corresponding to a center-of-mass~(c.m.s.)~energy of $\sqrt{S}=2\sqrt{E_p E_e}\approx 1.18 - 2.37$~TeV. Several studies have been performed to estimate the capability of such a machine to access the coupling of the Higgs boson to bottom quarks in a cleaner environment than at a hadron-hadron collider~\cite{UtaKlein09,Kuze09}. In particular, Han and Mellado~\cite{Han:2009pe} provided a detailed, cut-based leading-order (LO) analysis of the Higgs signal at the LHeC in the presence of various backgrounds. 

In addition to identifying the selection criteria most suitable for an extraction of the Higgs signal from backgrounds at an $ep$ collider, precise knowledge of the signal processes itself is crucial. We therefore provide accurate predictions for cross sections and kinematic distributions of Higgs production in association with a tagging jet at the LHeC, taking next-to-leading order (NLO) QCD corrections fully into account. We start with a brief outline of the calculational methods used in Sec.~\ref{sec:calc}. In Secs.~\ref{sec:num-cc} and \ref{sec:num-nc} we present our numerical results for the charged current  and neutral current  production modes, respectively.  Our focus is the estimate of the perturbative uncertainties associated with predictions for the signal process. Conclusions are given in Sec.~\ref{sec:conc}.

\section{Setup of the calculation}
\label{sec:calc}
At LO, Higgs production in $ep$ collisions mainly proceeds via lepton-quark scattering, mediated by the exchange of massive weak gauge bosons which in turn radiate off a Higgs boson. In the final state, the scattered quark gives rise to a jet~$j$.  Depending on the charge of the weak exchange boson, the respective production modes, $\ccp$ and $\ncp$,  are referred to as charged current (CC) and neutral current (NC), respectively. For simplicity, we will focus on the NC mode for the discussion of the technical details of the calculation. The treatment of the CC processes proceeds on the very same lines.

\begin{figure}
\bec
  \includegraphics[width=0.4\textwidth,clip]{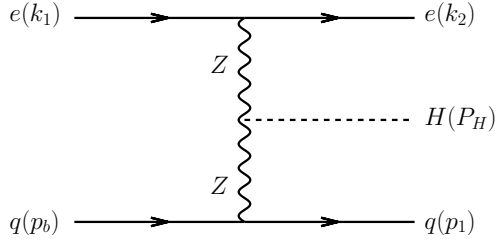}
  \caption{Leading order Feynman diagram for Higgs production at a lepton-hadron collider in the neutral current mode $\nc$.}
  \label{fig:born}
\eec
\end{figure}

Figure~\ref{fig:born} 
depicts the single Feynman diagram for the dominant NC production mode at parton level, 
\bea
e(k_1)\, q(p_b) & \to & e(k_2) \, q(p_1) \, H(P_H)\,.
\eea
The anti-quark initiated process, $e\bar q\to e \bar q H$, is easily obtained thereof by crossing relations. The scattering amplitudes, $\mc{M}^q_\mr{LO}$ and  $\mc{M}^{\bar q}_\mr{LO}$, for each partonic subprocess are computed numerically with the helicity-amplitude techniques of \cite{Hagiwara:1985yu,Hagiwara:1988pp}. 
For the hadronic jet cross section, each partonic matrix element squared,   $|\mc{M}^{b}_\mr{LO}|^2$, has to be convoluted with the parton distribution function of parton $b$ in the proton, $f_b(x_b,\muf)$, multiplied with the flux factor and a suitable jet-defining function, and integrated over the available phase space,  
\beq
\sigma^\mr{LO}(\ncp) = \sum_{b=q,\bar q}\int_0^1 dx_b f_q(x_b,\muf)\, \frac{1}{2\hat{s}}\,d\Phi_3\,
|\mc{M}_\mr{LO}^b|^2 F_J^{(1)}(p_1)
\,.
\eeq
Here, $x_b$ is the fraction of the proton momentum taken by parton $b$ at a factorization scale $\muf$. The partonic c.m.s.~energy squared is given by $\hat{s} = (k_1+p_b)^2$, $d\Phi_m$ denotes the $m$-particle phase space of the system, and the function $F_J^{(n)}$ defines the jet algorithm for an $n$-parton final state. 

At NLO, virtual and real-emission corrections to the tree-level amplitudes have to be considered. Since at the perturbative order we consider the color-neutral lepton-beam particles are not affected by QCD corrections,
the structure of the radiative corrections is simple and resembles that for deeply inelastic lepton-nucleon scattering. Very similar contributions arise in $H\,jj$ production via WBF in hadron-hadron collisions and have been computed in Refs.~\cite{Figy:2003nv,Berger:2004pca,Ciccolini:2007ec}. For our work, we adapt the NLO-QCD corrections of Ref.~\cite{Figy:2003nv} to Higgs production at lepton-hadron colliders.

\begin{figure}
\bec
\includegraphics[width=0.4\textwidth,clip]{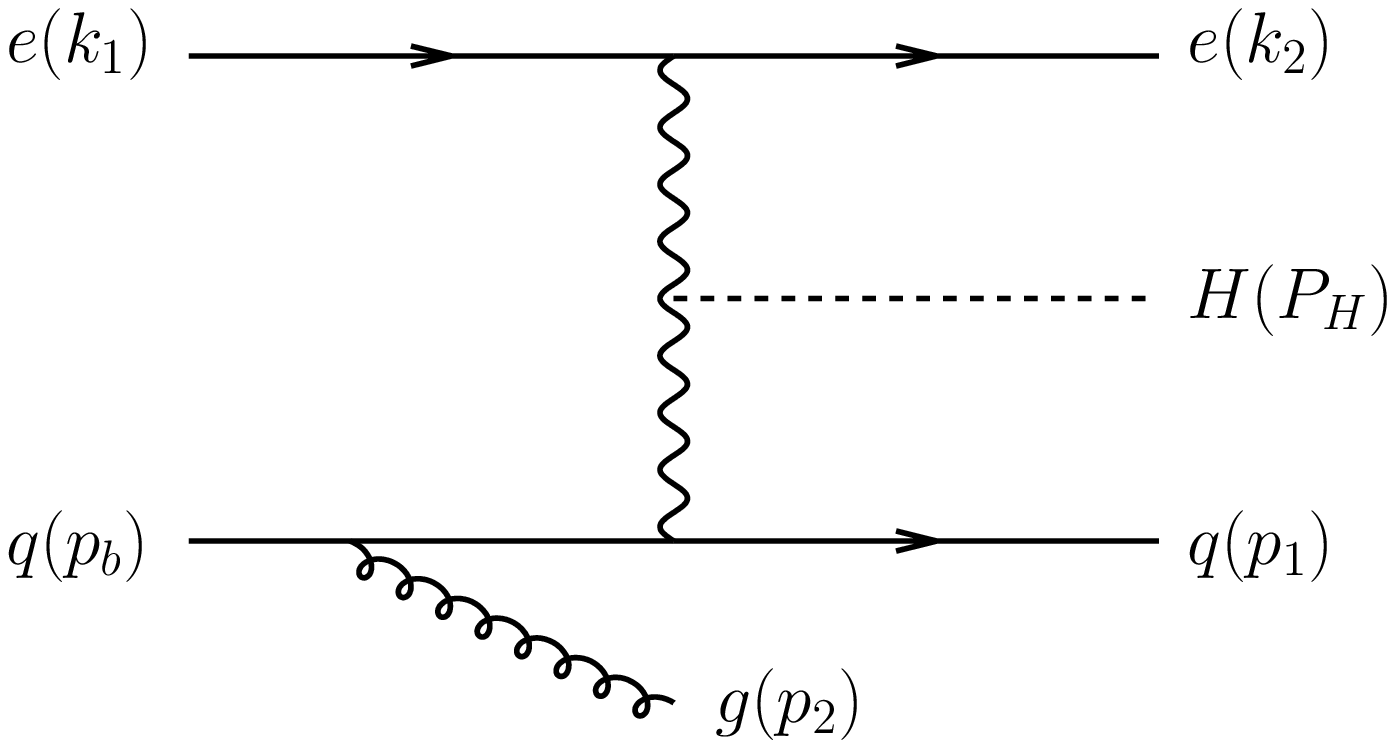}
\hs{1cm}
\includegraphics[width=0.4\textwidth,clip]{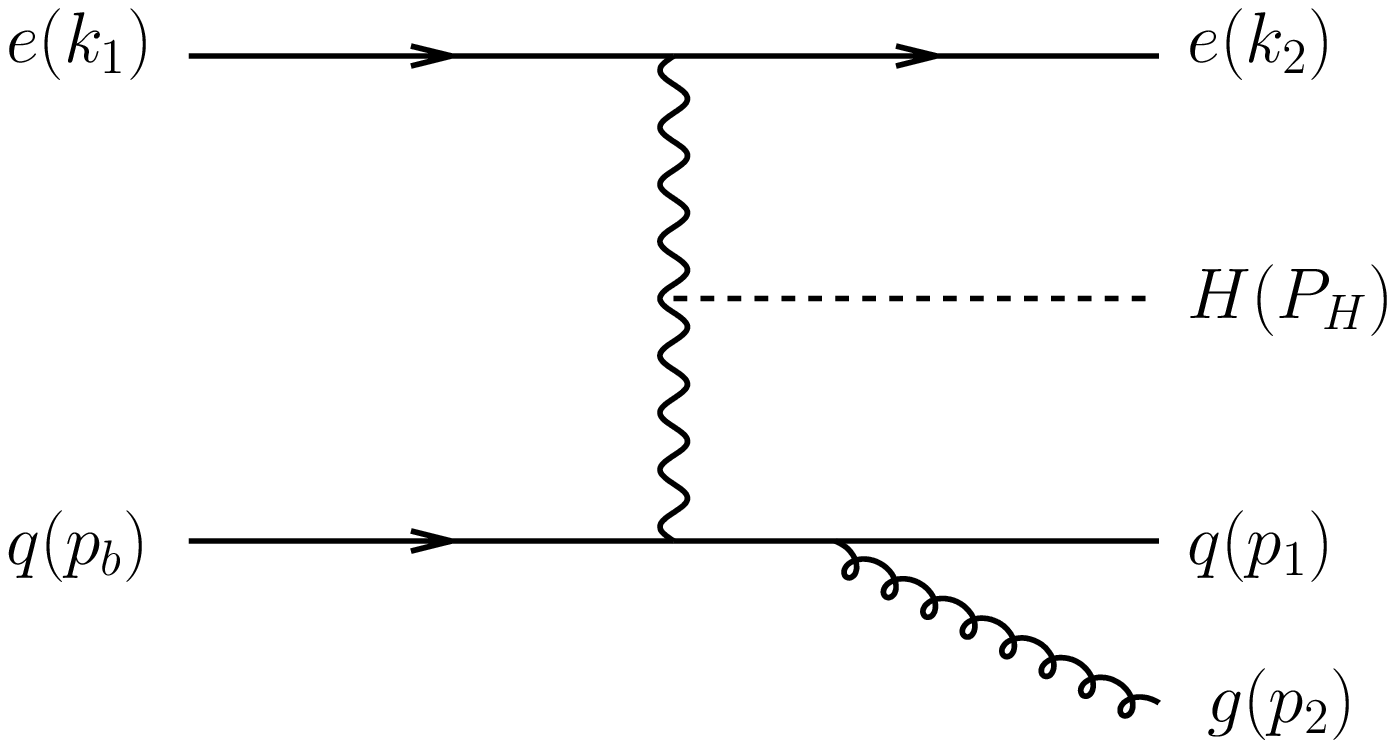}
\caption{Real emission contributions to Higgs production at a lepton-hadron collider in the neutral current mode $eq\to eqgH$.}
\label{fig:real}
\eec
\end{figure}

\begin{figure}
\bec
\includegraphics[width=0.4\textwidth,clip]{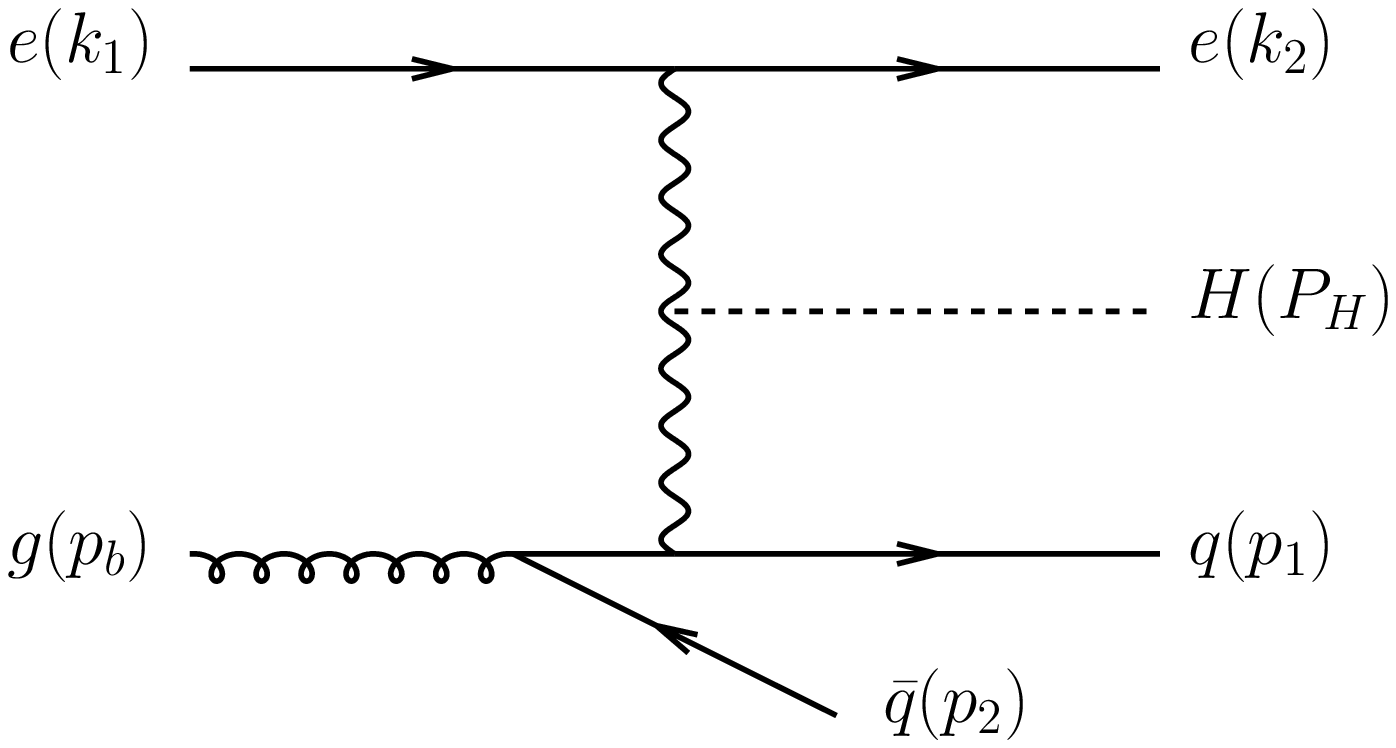}
\hs{1cm}
\includegraphics[width=0.4\textwidth,clip]{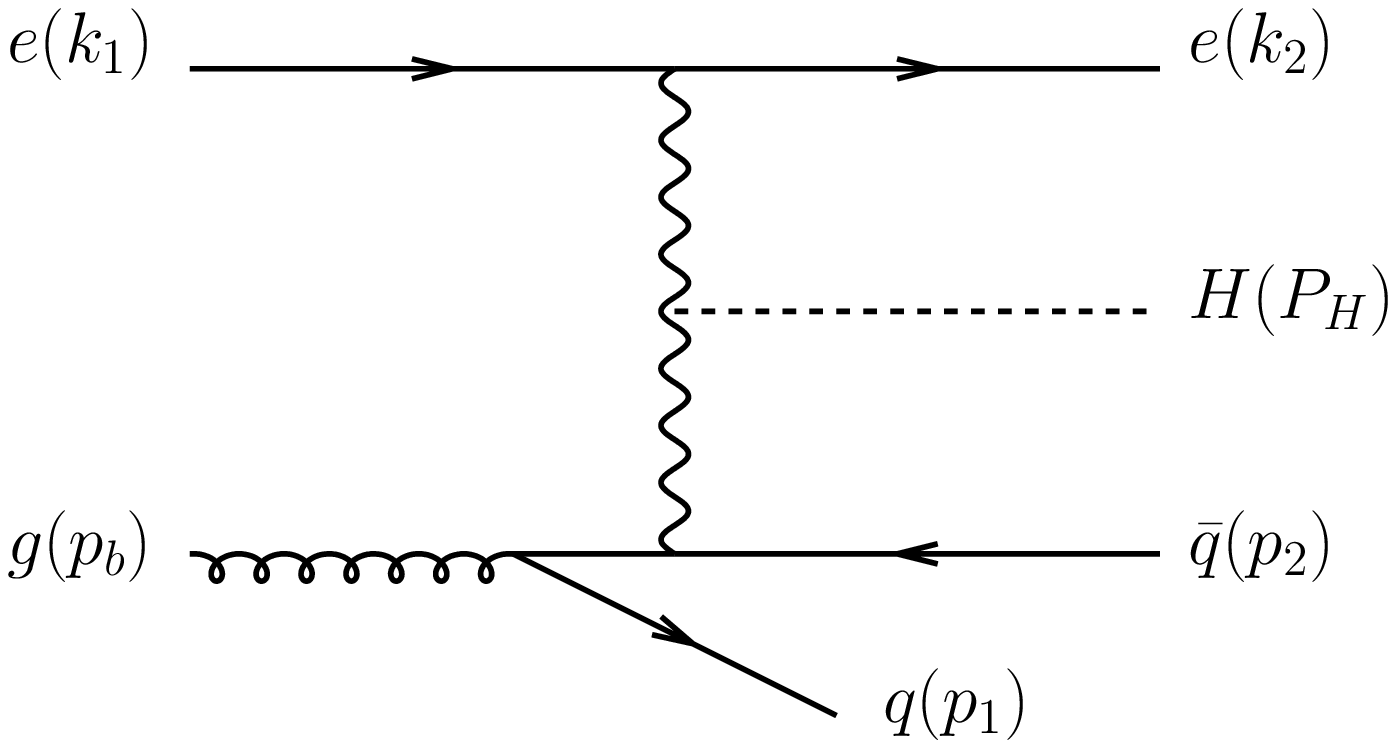}
\caption{Real emission contributions to Higgs production at a lepton-hadron collider in the neutral current mode $eg\to eq\bar qH$.}
\label{fig:realg}
\eec
\end{figure}

The real-emission amplitudes, $\mc{M}_R^b$, are obtained from the tree-level diagrams by attaching one gluon to the quark line in all possible ways.
The resulting diagrams are depicted for the representative quark-initiated subprocess, 
\beq
e(k_1) \, q(p_b) \to e(k_2)\, q(p_1)\, g(p_2) \, H(P_H)\,,
\eeq 
in Fig.~\ref{fig:real}.  
In addition to the $eq$ and $e\bar q$ channels with an extra gluon in the final state, now also the gluon-initiated subprocess,  
\beq
e(k_1) \, g(p_b) \to e(k_2)\, q(p_1)\, \bar q(p_2) \, H(P_H)\,,
\eeq 
has to be considered, cf.~Fig.~\ref{fig:realg}.  
For each channel, in regions of phase space where soft and/or collinear configurations can occur, singularities are encountered in the phase-space integrals of the real-emission amplitudes squared. These divergences are regularized dimensionally in $d= 4-2\eps$ space-time dimensions. The cancelation of the singularities is performed in terms of the dipole subtraction formalism of Ref.~\cite{Catani:1996vz}. With the radiative corrections being of the same form as in  WBF at hadron colliders, the relevant subtraction terms, $|\mc{M}_\mr{sing}^q|^2$, can be adapted from Ref.~\cite{Figy:2003nv} by replacing the entering Born-type parton-parton amplitudes with the appropriate lepton-parton contributions. The finite 2-parton cross section for the NC $eq$ subprocess is then of the form
\beq
\label{eq:sig-real}
\sigma_2^{\mr{NLO},q} = \int_0^1 dx_b f_q(x_b,\muf)\,\frac{1}{2\hat{s}}\,d\Phi_4\,
\Biggl\{ |\mc{M}_R^q|^2 F_J^{(2)}(p_1,p_2) - |\mc{M}_\mr{sing}^q|^2 F_J^{(1)}(\tilde{p}_1)
\Biggr\}\,,
\eeq
where, in the notation of \cite{Catani:1996vz}, $\tilde{p}_1$ corresponds to a set of parton momenta approaching Born kinematics in the collinear and soft limits, $\tilde{p}_1\to p_1$. In these regions, the jet-defining function has to obey $F_J^{(2)}\to F_J^{(1)}$. 
Expressions similar to Eq.~(\ref{eq:sig-real}) are obtained for the anti-quark- and the gluon-initiated subprocesses. 

\begin{figure}
\bec
 \includegraphics[width=0.4\textwidth,clip]{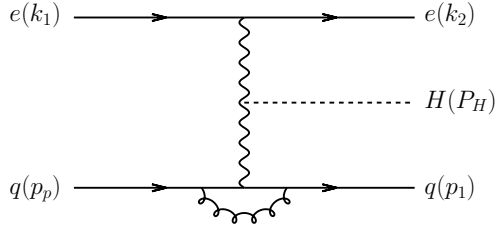}
  \caption{Virtual contribution to Higgs production at a lepton-hadron collider in the neutral current mode $\nc$.}
  \label{fig:virt}
\eec
\end{figure}

In order to arrive at well-defined results for physical observables, the counter-terms that have been introduced in Eq.~(\ref{eq:sig-real}) are integrated  analytically in $d$ dimensions over the phase-space of the potentially soft or collinear parton, yielding  
\beq
\langle \mathbf{I}(\eps)\rangle  = 
|\mc{M}_\mr{LO}^{\,q}|^2\,
\frac{\alpha_s (\mur)}{2\pi} \,C_F\,
\Biggl(\frac{4\pi\mur^2}{Q^2}\Biggr)^{\eps}\,
\Gamma(1+\eps)\,
\Biggl[
\frac{2}{\eps^2}+\frac{3}{\eps} + 9 -\frac{4}{3}\pi^2
\Biggr]\,,
\eeq
where $\alpha_s(\mur)$ is the strong coupling as a function of the renormalization scale $\mur$, $C_F = 4/3$, and $Q^2 = -(p_b-p_1-p_2)^2$ denotes the virtuality of the weak boson attached to the quark line.   
The virtual corrections comprise vertex corrections to the quark line as depicted in Fig.~\ref{fig:virt}. 
They are entirely proportional to the Born amplitude such that the relevant interference with the tree-level amplitude is given by  
\beq
2 \mr{Re}\,[\mc{M}^q_V\mc{M}^{\,q,\star}_{\mr{LO}}] = 
|\mc{M}_\mr{LO}^{\,q}|^2\,
\frac{\alpha_s (\mur)}{2\pi} \,C_F\,
\Biggl(\frac{4\pi\mur^2}{Q^2}\Biggr)^{\eps}\,
\Gamma(1+\eps)\,
\Biggl[
-\frac{2}{\eps^2}-\frac{3}{\eps} + c_{\mr{virt}}
\Biggr]\,,
\eeq
where $c_\mr{virt}$ stands for a constant given by  $c_\mr{virt} = \pi^2/3-7$ in dimensional reduction ($c_\mr{virt} = \pi^2/3-8$ in dimensional regularization). 
After adding the integrated counter-terms, the resulting 1-parton contribution to the NLO cross section is finite and for the NC $eq$ subprocess takes the form 
\bea
\sigma_1^{\mr{NLO},q} &=& \int_0^1 dx_b f_q(x_b,\muf)\, \frac{1}{2\hat{s}}\,d\Phi_3\,
|\mc{M}_\mr{LO}^q|^2 \, F_J^{(1)}(p_1)
\Biggl[ 1 + \frac{\alpha_s (\mur)}{2\pi} \,C_F
\Biggl(
9-\frac{4}{3}\pi^2 + c_{\mr{virt}}
\Biggr)
\Biggr].
\eea

A remaining divergent piece of the integrated counter terms is factorized into the distribution function of the incoming parton. For the quark-initiated contribution, the surviving finite collinear term is given by 
\bea
\label{eq:coll}
\sigma_{1,\mr{coll}}^{\mr{NLO},q}&=& 
\int_0^1 dx_b f_q^c(x_b,\muf, \mur)\, \frac{1}{2\hat{s}}\,d\Phi_3\,
|\mc{M}_\mr{LO}^q|^2 F_J^{(1)}(p_1)\,,
\eea
where $f_q^c(x_b,\muf, \mur)$ is defined by 
\bea      
f_{q}^c(x,\muf,\mur)&=& \frac{\alpha_s(\mur)}{2\pi} 
\int_x^1 \frac{dz}{z}
\lg f_{g} \(\frac{x}{z},\muf\) A(z)\right.
\nonumber \\ 
&& + \left.
\lq f_{q} \(\frac{x}{z},\muf\)-z f_{q} \(x,\muf\) \rq B(z)
+f_{q}\(\frac{x}{z},\muf\) C(z) \rg
\nonumber \\ 
&& + \frac{\alpha_s(\mur)}{2\pi} f_{q} (x,\muf) D(x)\;,
\eea
with the integration kernels
\bea   
A(z) &=& T_F\lq z^2+(1-z)^2\rq\ln\frac{Q^2(1-z)}{\muf^2 z}
     +2T_F\; z(1-z)\;, \\
B(z) &=& C_F\lq
\frac{2}{1-z} \ln\frac{Q^2(1-z)}{\muf^2} -\frac{3}{2}\frac{1}{1-z}
\rq\;, \\
C(z) &=& C_F\lq 1-z-\frac{2}{1-z}\ln z - 
(1+z) \ln\frac{Q^2(1-z)}{\muf^2 z} \rq\;, \\
D(x) &=& C_F \lq \frac{3}{2} \ln\frac{Q^2}{\muf^2 (1-x)}
+2\ln(1-x)\ln\frac{Q^2}{\muf^2} +\ln^2(1-x) 
+ \pi^2 -\frac{27}{2} -c_{\rm virt} \rq \,, \phantom{aa}
\eea
where $T_F= 1/2$. Analogous expressions are encountered for the anti-quark initiated subprocesses. 

The NC cross-section contributions discussed above and the respective CC expressions have been implemented in a flexible Monte-Carlo program that allows for the calculation of experimentally accessible observables within realistic acceptance cuts at NLO-QCD accuracy. In order to ensure the reliability of our calculation, a number of tests has been performed:
\begin{itemize}
\item
The tree-level and the real-emission amplitudes have been compared to the corresponding expressions generated automatically by {\tt MadGraph}~\cite{Stelzer:1994ta} for a representative set of phase-space points. We found full agreement.
Since the virtual amplitudes are just multiples of the Born cross section, they do not need to be tested separately. 

\item
The integrated LO cross sections have been compared to the corresponding results of the {\tt MadEvent} package~\cite{Maltoni:2002qb,Alwall:2007st} 
for the inclusive selection cuts to be described in Sec.~\ref{sec:num}. 
The cross sections agree within the numerical accuracy of the two programs. 

\item
The real-emission matrix elements have been found to vanish when the polarization vector of the external gluon is replaced with its momentum. This procedure tests the QCD gauge invariance of the real-emission amplitudes.

\item
To probe our implementation of the dipole subtraction, we checked that the real-emission contributions approach the subtraction terms in singular regions of phase space. 

\end{itemize}

\section{Numerical results}
\label{sec:num}
In order to assess the impact of NLO-QCD corrections on cross sections and kinematic distributions, we consider Higgs production at the LHeC within different settings. We present numerical results applying only minimal  selection cuts first and then turn to the settings which have been suggested in Ref.~\cite{Han:2009pe} for extracting the Higgs signal from various backgrounds at the LHeC.
Unless stated otherwise, we assume electron-proton collisions with 
\beq
E_e = 140~\mr{GeV}\,\;\mr{and}\,\;
E_p = 7~\mr{TeV}\,,
\eeq
and a SM-like Higgs boson with a mass of $M_H = 120$~GeV. Quark and lepton masses are set to zero throughout, and contributions from incoming top and bottom quarks are neglected.  

To simulate a generic Higgs decay without specifying a particular channel, we generate an isotropic Higgs boson decay into two massless particles  (which represent $\gamma\gamma$ or $b\bar b$ final states), requiring each decay particle, labeled $d$, to be separated from a jet in the rapidity-azimuthal angle plane by 
\beq
\label{eq:rjd-cut}
\Delta R_{j{\dec}}>0.4\,. 
\eeq
The respective branching ratio 
$BR(H\to\dec\dec)$ is not included in the numerical results we present below. 

Depending on the order in perturbation theory at which an observable is being computed, we use the LO or the NLO set of the MSTW08 parton distribution functions (PDFs) \cite{Martin:2009iq}. For the strong coupling, which enters only at NLO, we take the  appropriate two-loop expression. 
We have chosen the weak gauge boson masses, $M_W=80.423$~GeV, $M_Z=91.188$~GeV, and the Fermi constant, $G_F=1.166\times~10^{-5}/$~GeV$^2$, as electroweak input parameters. Thereof, $\alpha_\mr{QED}$ and $\sin^2\theta_W$ are computed via LO electroweak relations. 
Final-state partons are recombined into jets according to the $k_T$~algorithm \cite{Catani:1992zp,Catani:1993hr,Ellis:1993tq,Blazey:2000qt} with a resolution parameter $D=0.4$. Jets are required to have a  transverse momentum of
\beq
\label{eq:ptjet-cut}
p_T^\mr{jet} > 15~\mr{GeV}\,,
\eeq
and events which do not exhibit at least one jet are disregarded. 
In configurations with more than one identified jet, the jet of highest transverse momentum is referred to as ``tagging jet''. Below, we will refer to observables within the cuts of Eqs.~(\ref{eq:rjd-cut}) and (\ref{eq:ptjet-cut}) as ``inclusive''. 

For clarity, we will discuss the CC and NC production modes separately in the following. 

%
%
\subsection{Charged current processes}
\label{sec:num-cc}
CC processes represent the major source of Higgs bosons in $ep$ collisions due to the value of the charged weak boson coupling to the electron, which is much larger than the respective neutral current coupling. 
At LO, the signal topology consists of large missing energy from the neutrino, a $\dec\dec$ pair from the Higgs decay, and a forward energetic tagging jet. At NLO, this jet may be composed of two partons, or we may encounter a second, well-separated jet. 

\begin{figure}[!tb]
\bec
\includegraphics[width=0.4\textwidth,clip]{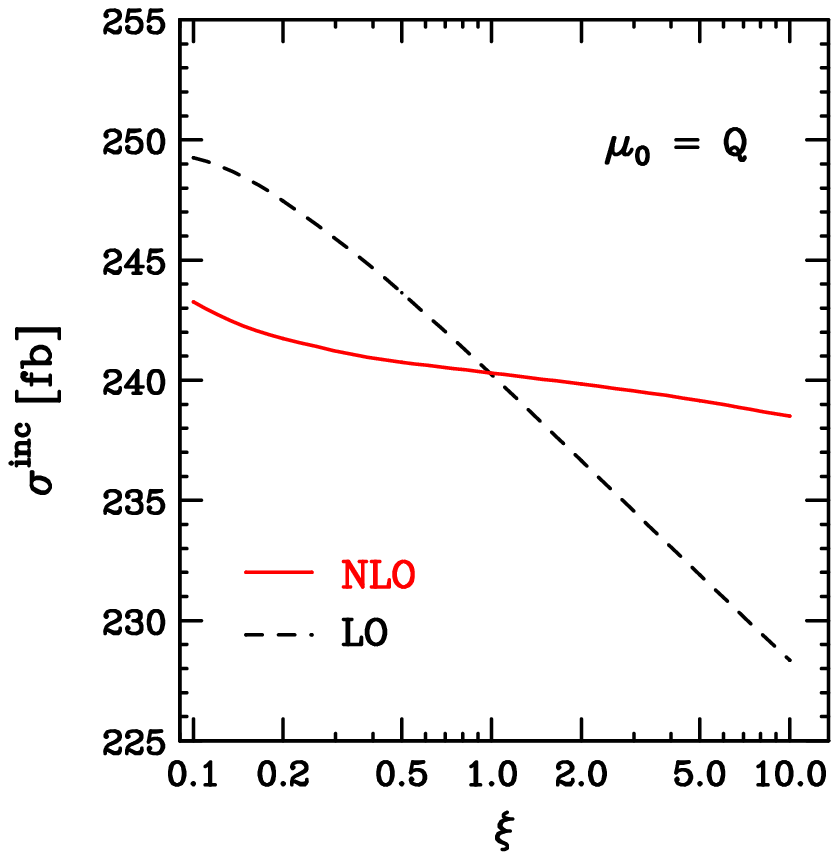}
\caption{
Dependence of the $\ccp$ cross section, $\sigma_\mr{CC}^\mr{inc}$, at the LHeC on the factorization and renormalization scales. The NLO curves show $\sigma_\mr{CC}^\mr{inc}$ as a function of the scale parameter $\xi$ for $\mur=\muf=\xi\mu_0$ (solid red). The LO cross section only depends on $\muf=\xi\mu_0$ (dashed black). }
\label{fig:cc-scale-dep}
\eec
\end{figure}

In Fig.~\ref{fig:cc-scale-dep},  
the integrated CC cross section within the cuts specified in  Eqs.~(\ref{eq:rjd-cut})~and~(\ref{eq:ptjet-cut}), $\sigma_\mr{CC}^\mr{inc}$, 
is displayed as a function of the scale parameter $\xi$, which is related to the factorization and renormalization scales, $\muf$ and $\mur$, via
\beq
\muf = \xi \, \mu_0\;\;\mr{and}\;\;
\mur = \xi \, \mu_0\;\;\mr{with}\;\;\mu_0 = Q\,. 
\label{eq:scale}
\eeq
The LO cross section depends only on $\muf$, and the variation of $\sigma_\mr{CC}^\mr{inc}$ with $\xi$ resembles the scale dependence of the quark distribution function $f_q(x,\muf)$ at relatively large values of $x$. At NLO, the $\muf$-dependence of the parton distributions is partly canceled by respective contributions in the partonic cross section. The renormalization scale enters via the strong coupling $\alpha_s(\mur)$. Due to the small size of the NLO-QCD corrections, the overall dependence on $\mur$ is small. 
Qualitatively, the improvement of the scale dependence when going from LO to NLO is very similar to what has been observed for WBF processes in hadronic collisions \cite{Jager:2006zc,Jager:2006cp,Bozzi:2007ur,Jager:2009xx}. It is interesting to note, however, that the relative size of the scale uncertainties is much smaller in the $ep$ case, due to the presence of only one parton distribution function as opposed to two in $pp$ collisions. 

Additional uncertainties could arise from the incomplete knowledge of the parton distribution functions of the proton. In order to estimate this uncertainty, we have calculated $\sigma_\mr{CC}^\mr{inc}$ for two scale choices, $\muf=\mur=Q$ and $\muf=\mur=M_W$, using two different PDF parameterizations: our default set MSTW08 and the CTEQ6 set \cite{Pumplin:2002vw}. In particular, the CTEQ6M set is used at NLO, and the CTEQ6L1 set at LO. Our results are listed in Table~\ref{tab:sig-pdf}. 
%
%
\begin{table}[!h]
\begin{center}
\begin{tabular}{|c|c|c|c|c|}
\hline
PDF set  &$\sigma_\mr{CC}^\mr{inc,LO}(\mu_0=M_W)$&$\sigma_\mr{CC}^\mr{inc,NLO}(\mu_0=M_W)$
	 &$\sigma_\mr{CC}^\mr{inc,LO}(\mu_0=Q)$  &$\sigma_\mr{CC}^\mr{inc,NLO}(\mu_0=Q)$  
	 \\
	 \hline
CTEQ6 &$241.83$~fb&$240.82$~fb&$241.89$~fb&$239.78$~fb 
	\\
	\hline
MSTW08 &$240.11$~fb&$241.35$~fb&$240.24$~fb&$240.30$~fb
	\\
	\hline
\end{tabular}
\caption{
\label{tab:sig-pdf}
Inclusive cross section, $\sigma_\mr{CC}^\mr{inc}$, for $\ccp$ at the LHeC for the scale choices $\mu_0=M_W$ and $\mu_0=Q$ with $\xi=1$, and for two different sets of parton distribution functions. The statistical errors of the quoted results are at
the sub-permille level and therefore not given explicitly.  
}	
\vspace*{-.5cm}
\end{center}
\end{table}
%
Very similar results are obtained when $M_W$ is used as a default scale rather than $Q$, since $Q$ tends to be close to the weak boson mass scale in this class of reactions. The numerical uncertainty of $\sigma_\mr{CC}^\mr{inc}$ due to the PDF parameterization is small both at LO and NLO, with relative differences between the two choices being at the permille level. In the following we will stick to the MSTW08 parton distributions and set $\muf=\mur = Q$, unless stated otherwise. 

\begin{figure}[!tp]
\bec
\includegraphics[width=0.9\textwidth,clip]{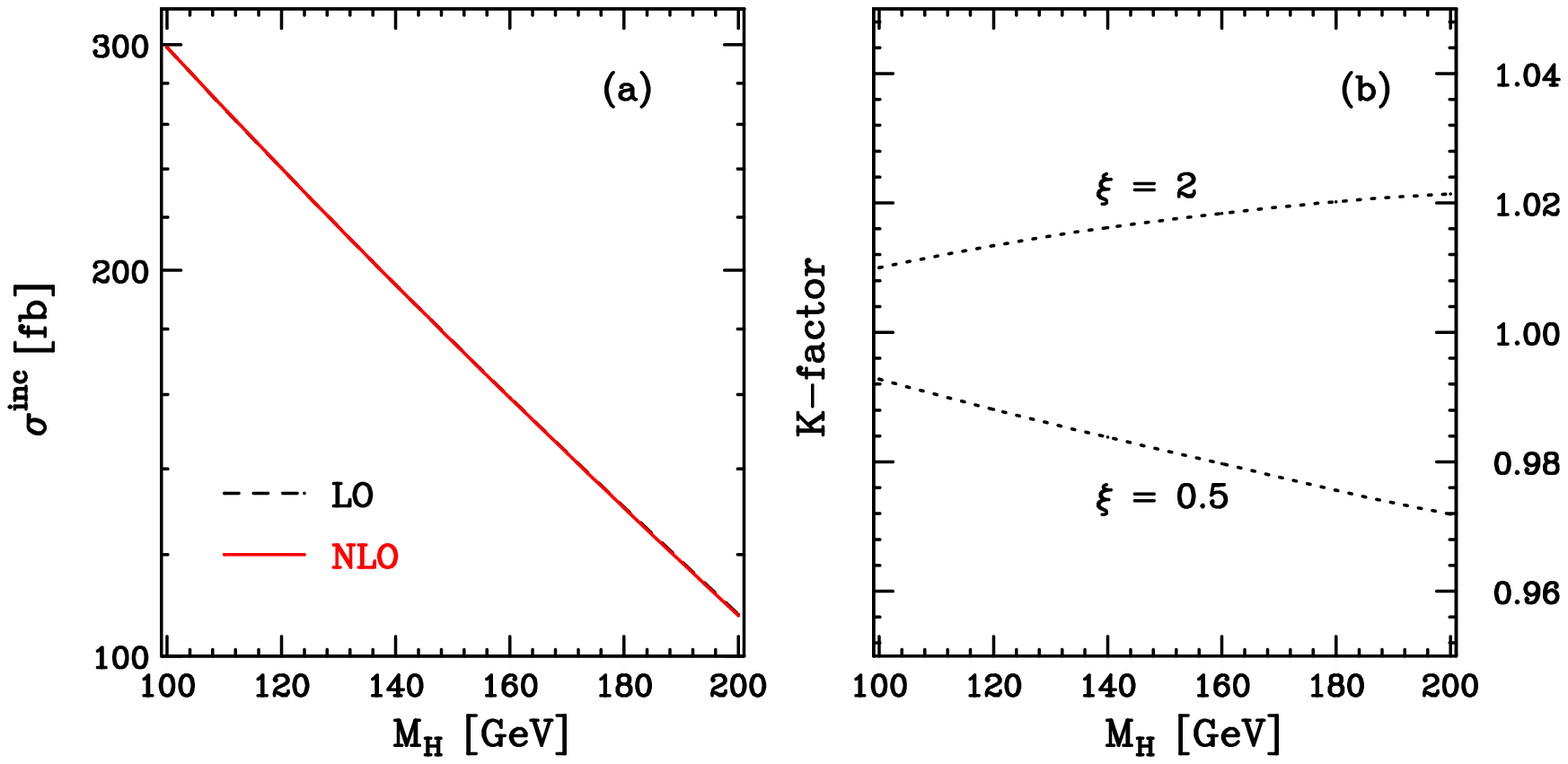}
\caption{
Panel~(a) shows the dependence of the $\ccp$ cross section, $\sigma_\mr{CC}^\mr{inc}$, at the LHeC on the mass of the Higgs boson at LO (black dashed) and NLO (red solid) for $E_e=140$~GeV. In panel~(b) the corresponding $K$-factor as defined in Eq.~(\ref{eq:kfac}) is shown for different values of the factorization and renormalization scales, 
$\mur=\muf=\xi\,Q$.   }
\label{fig:cc-mh-dep}
\eec
\end{figure}

\begin{figure}[!tp]
\bec
\includegraphics[width=0.9\textwidth,clip]{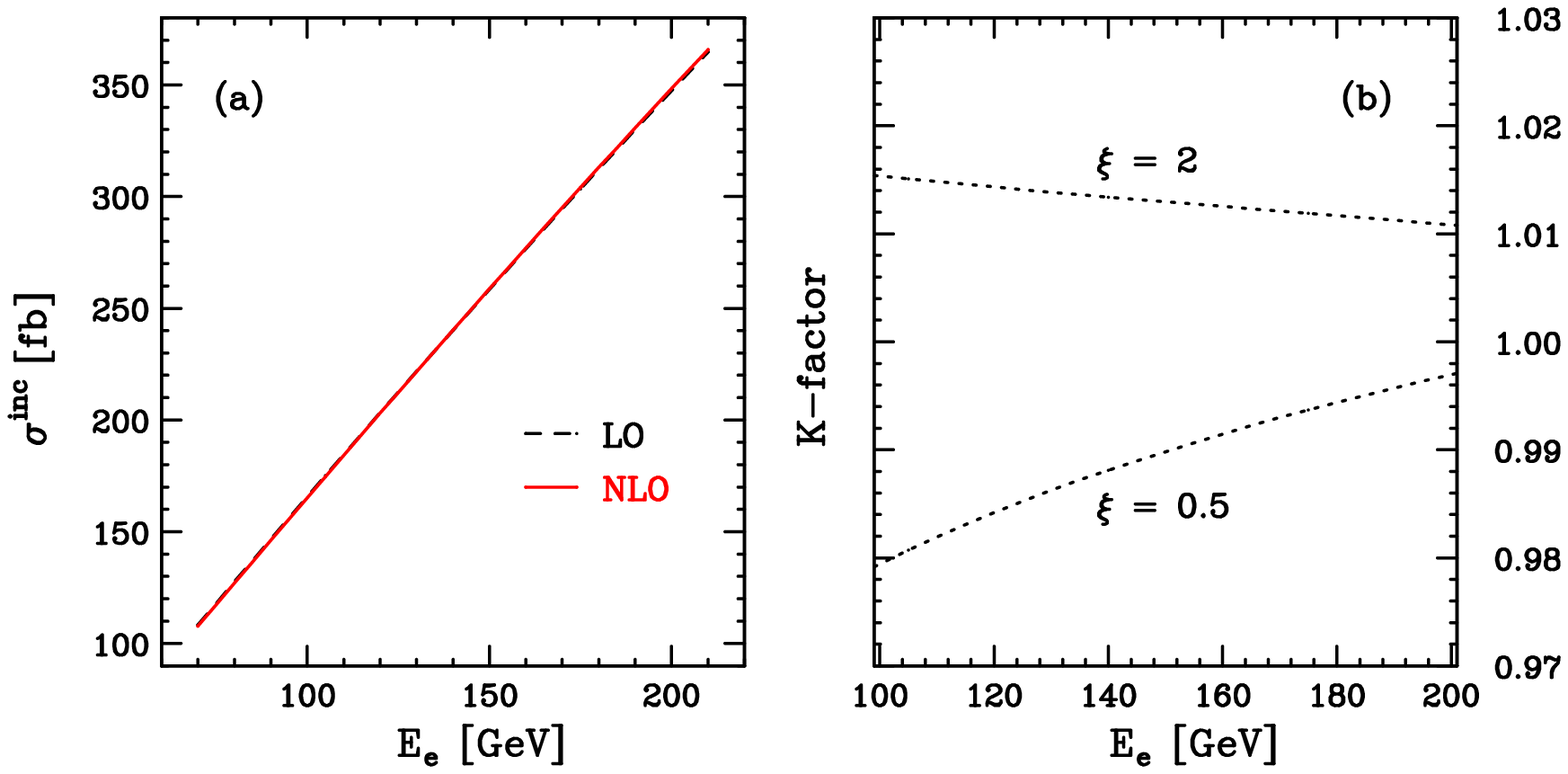}
\caption{
Panel~(a) shows the dependence of the $\ccp$ cross section, $\sigma_\mr{CC}^\mr{inc}$, at the LHeC on the energy of the electron beam at LO (black dashed) and NLO (red solid) for $M_H=120$~GeV. In panel~(b) the corresponding $K$-factor as defined in Eq.~(\ref{eq:kfac}) is shown for different values of the factorization and renormalization scales, 
$\mur=\muf=\xi\,Q$.    }
\label{fig:cc-elep-dep}
\eec
\end{figure}

In order to estimate the impact of the NLO-QCD corrections on various observables, we introduce the dynamical $K$-factor of a distribution $d\sigma/d\mc{O}$, 
\beq
K(x) = \frac{d\sigma^\mr{NLO}(\mur,\muf)/d\mc{O}}{d\sigma^\mr{LO}(\muf)/d\mc{O}}\,.
\label{eq:kfac}
\eeq
In Fig.~\ref{fig:cc-mh-dep}, 
$\sigma_\mr{CC}^\mr{inc}$ is shown as a function of the Higgs boson mass  together with the associated $K$-factor for two different choices of the scale parameter $\xi$ with $\mu_0=Q$. The NLO-QCD corrections are small over the entire mass range considered with relative scale uncertainties increasing with $M_H$. 
Fig.~\ref{fig:cc-elep-dep} 
illustrates that the scale uncertainties go down when the energy of the electron beam is increased. Considering this feature together with results on the signal significance reported in Ref.~\cite{Han:2009pe}, we conclude that for a given Higgs mass large values of $E_e$ are favorable for a clean extraction of the Higgs signal at the LHeC. In the following, we will stick to our default choice, $E_e = 140$~GeV and $M_H=120$~GeV.

Having estimated theoretical uncertainties for $\sigma_\mr{CC}^\mr{inc}$ we now turn to a study of CC Higgs production at the LHeC with selection cuts for an optimal signal significance, following  Ref.~\cite{Han:2009pe}. 
For each decay product $d$ of the Higgs boson we require
\beq
\label{eq:cc-dec-cuts}
p_T^{\dec}> 30~\mr{GeV}\,,\;\;
|\eta^{\dec}|<2.5\,,\;\;
\Delta R_{j{\dec}}>0.4\,,
\eeq
with $p_T^{\dec}$ denoting its transverse momentum and $\eta^{\dec}$ its pseudo-rapidity being given in the laboratory frame of the $ep$ system. 
The neutrino gives rise to missing energy in the detector, which we request to obey 
\beq
E_T^\mr{miss}> 25~\mr{GeV}\,. 
\eeq
Any identified jet has to fulfill 
\beq
p_T^\mr{jet} > 15~\mr{GeV}\;\;\mr{and}\;\;
|y^\mr{jet}|<5\,,
\eeq
where $y^\mr{jet}$ denotes the rapidity of a jet. 
For the tagging jet we furthermore demand
\beq
p_T^\mr{tag} > 30~\mr{GeV}\;\;\mr{and}\;\;
-5<y^\mr{tag}<-1\,.
\eeq
Note that (contrary to Ref.~\cite{Han:2009pe}) we count positive rapidity in the direction of the electron beam. The invariant mass of the Higgs boson candidate and the tagging jet, $M_\mr{H,tag}$, is required to be large, 
\beq
\label{eq:cc-mhj-cut}
M_\mr{H,tag} > 250~\mr{GeV}\,.
\eeq
The CC cross section within the cuts given in  Eqs.~(\ref{eq:cc-dec-cuts})--(\ref{eq:cc-mhj-cut}), $\sigma_\mr{CC}^\mr{cuts}$, changes by about 6.5\% at LO and 8$\permil$ at NLO when $\mur$ and $\muf$ are varied simultaneously in the range $Q/2$ to $2 Q$. The scale dependence of the CC cross section is thus slightly larger after the application of selection cuts than in the inclusive case, but still very small. In particular, the NLO prediction is extremely stable with respect to scale variations irrespective of the selection cuts applied, which indicates that Higgs production at the LHeC is under excellent control perturbatively.  

\begin{figure}[!tp]
\bec
\includegraphics[width=0.9\textwidth,clip]{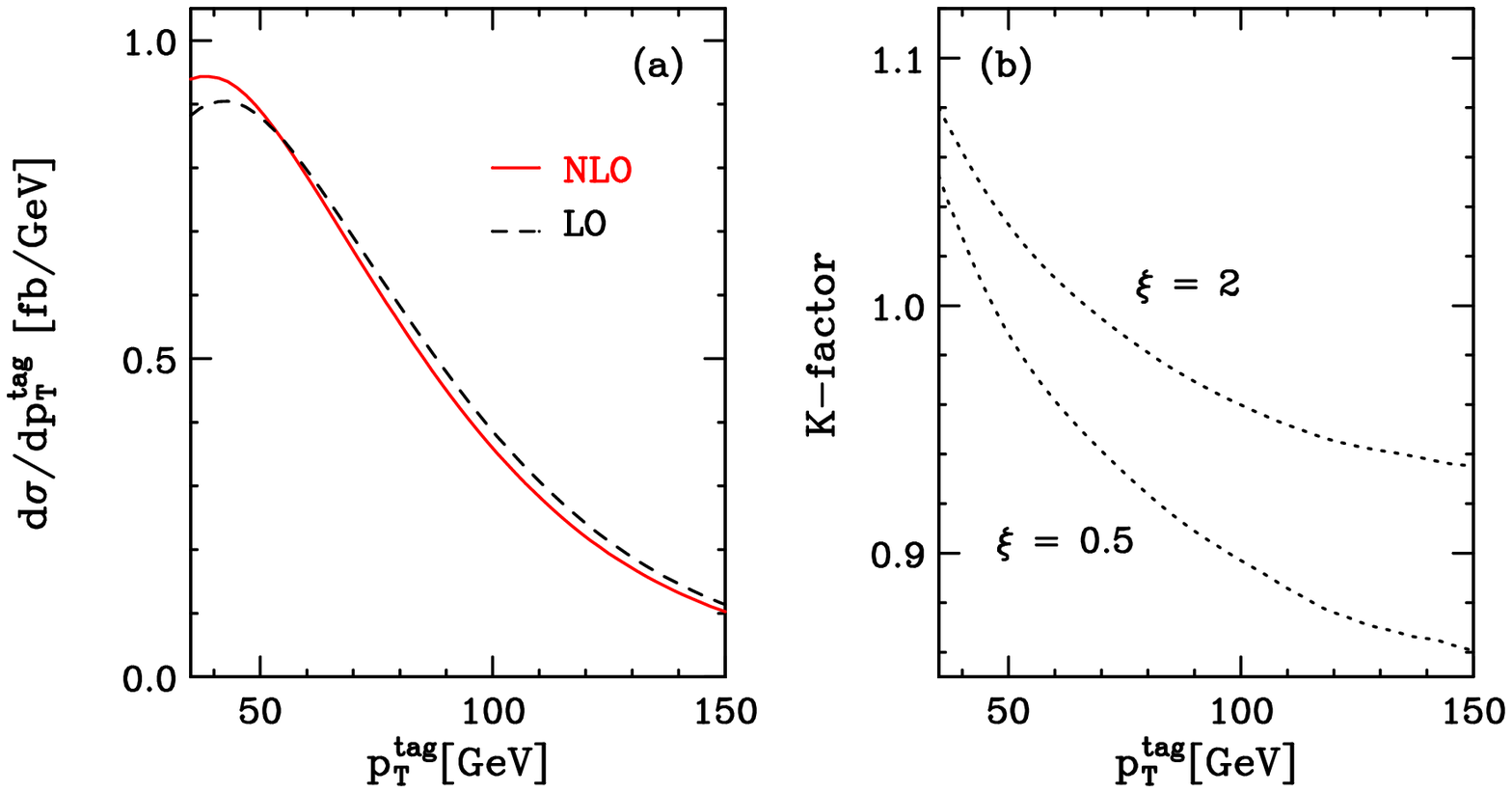}
\caption{
Transverse momentum distribution of the tagging jet in $\ccp$ at the LHeC at LO (dashed black) and NLO (solid red) [panel~(a)] and associated $K$-factor [panel~(b)] as defined in Eq.~(\ref{eq:kfac}) for different values of the factorization and renormalization scales, 
$\mur=\muf=\xi\,Q$.     }
\label{fig:cc-pt-tag}
\eec
\end{figure}

Similarly to WBF processes at hadron colliders, Higgs production processes at the LHeC feature tagging jets with pronounced kinematic properties that help to extract the signal from a variety of backgrounds. In $\ccp$, the tagging jet tends to be located in the far-backward region of the detector, along the direction of the incoming proton, and can be well separated from the decay products of the Higgs boson. This is particularly important, if Higgs production at the LHeC is to be utilized for an extraction of the coupling of the Higgs boson to bottom quarks. In this context, it is important to estimate the impact of NLO-QCD corrections on kinematic distributions of the tagging jet, such as its transverse momentum distribution displayed in Fig.~\ref{fig:cc-pt-tag} 
for $\mur=\muf=Q$ together with the associated $K$-factor for two different values of the factorization and renormalization scales. 
The corrections to $d\sigma/dp_T^\mr{tag}$ are slightly positive at low $p_T^\mr{tag}$, but increasingly negative towards larger transverse momenta. The difference between the $K$-factors for the two considered values of $\xi$ indicates the scale uncertainty of the distribution, which also increases with $p_T^\mr{tag}$.

%
%
\subsection{Neutral current processes}
\label{sec:num-nc}
Cross sections for NC Higgs production via  $\ncp$ are considerably smaller than  the related CC quantities. 
Due to the charged lepton in the final state, however, the event reconstruction in the NC mode is superior to the CC processes with a neutrino that can only be identified indirectly as missing energy in the detector. Moreover, NC processes yield information on the coupling of the Higgs boson to neutral gauge bosons rather than charged ones which are probed in CC Higgs production at the LHeC. 

\begin{figure}[!tp]
\bec
\includegraphics[width=0.4\textwidth,clip]{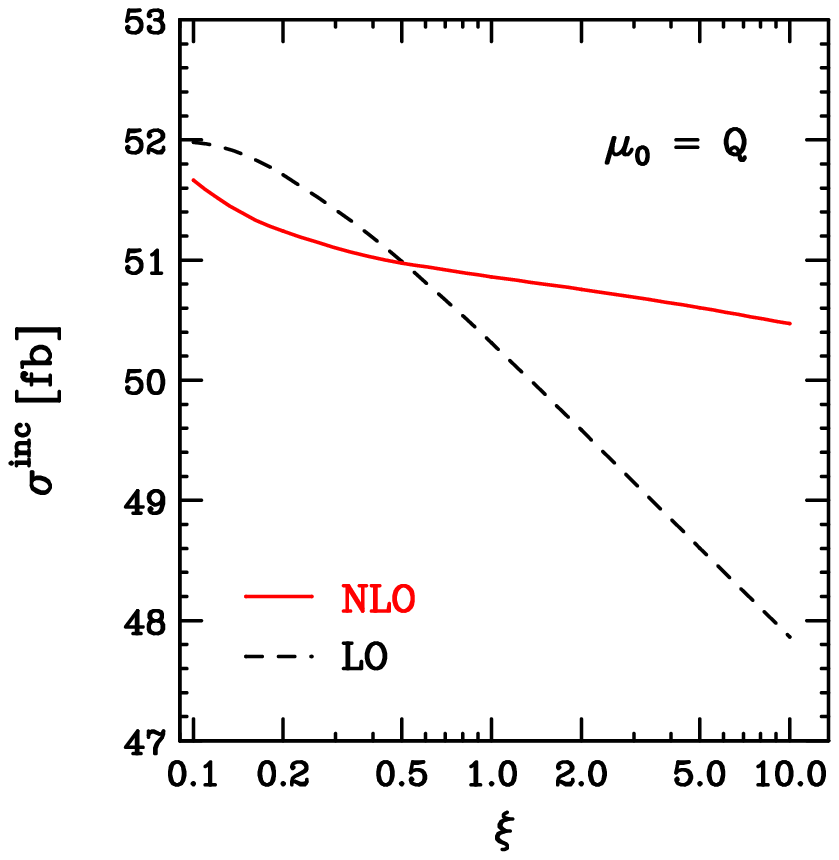}
\caption{
Dependence of the $\ncp$ cross section, $\sigma_\mr{NC}^\mr{inc}$, at the LHeC on the factorization and renormalization scales. The NLO curves show $\sigma_\mr{NC}^\mr{inc}$ as a function of the scale parameter $\xi$ for $\mur=\muf=\xi \mu_0$ (solid red). The LO cross section only depends on $\muf=\xi \mu_0$ (dashed black). }
\label{fig:nc-scale-dep}
\eec
\end{figure}

\begin{figure}[!tp]
\bec
\includegraphics[width=0.9\textwidth,clip]{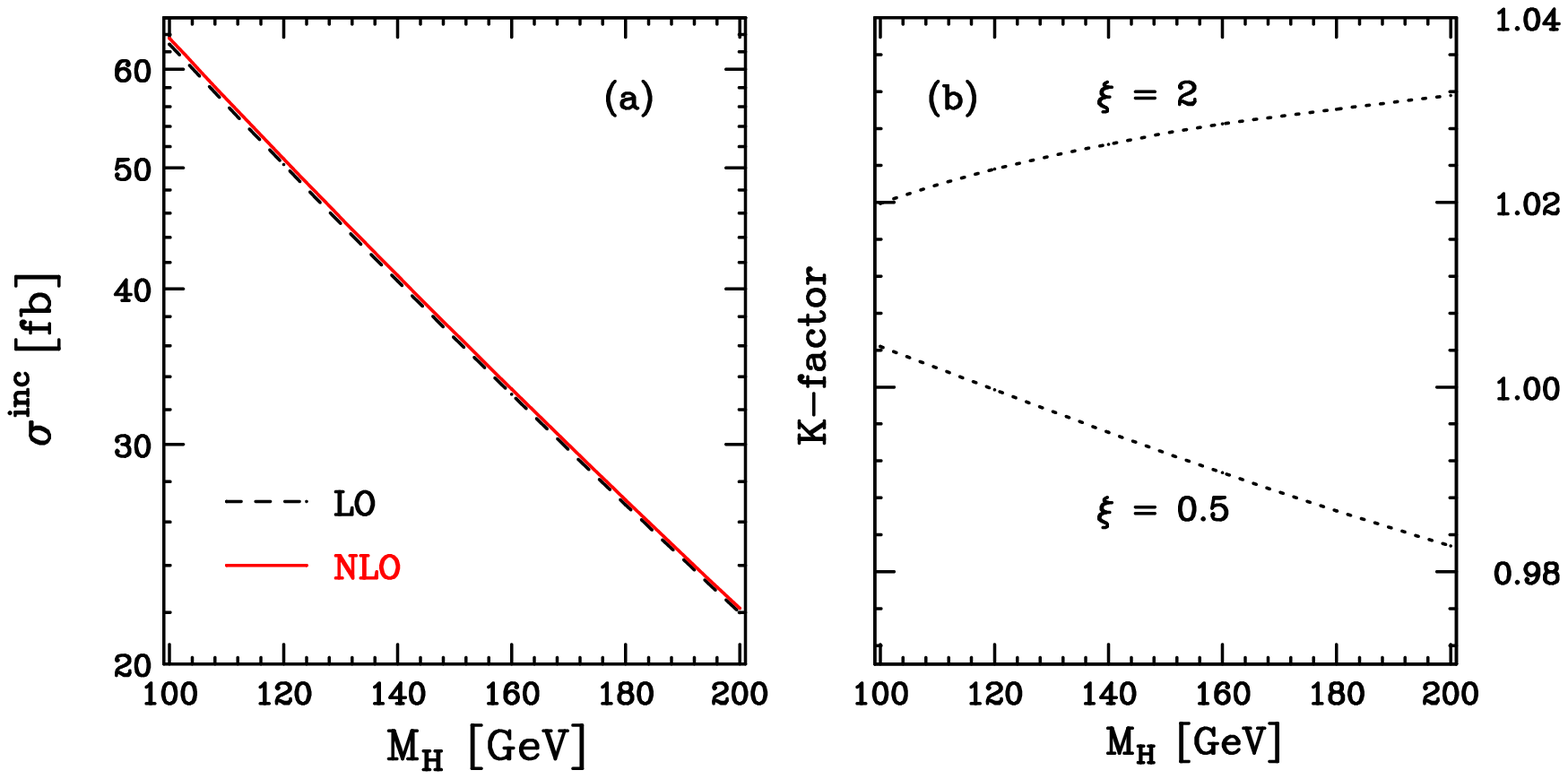}
\caption{
Panel~(a) shows the dependence of the $\ncp$ cross section, $\sigma_\mr{NC}^\mr{inc}$, at the LHeC on the mass of the Higgs boson at LO (black dashed) and NLO (red solid) for $E_e=140$~GeV.  In panel~(b) the corresponding $K$-factor as defined in Eq.~(\ref{eq:kfac}) is shown for different values of the factorization and renormalization scales, 
$\mur=\muf=\xi\,Q$.   }
\label{fig:nc-mh-dep}
\eec
\end{figure}

\begin{figure}[!tp]
\bec
\includegraphics[width=0.9\textwidth,clip]{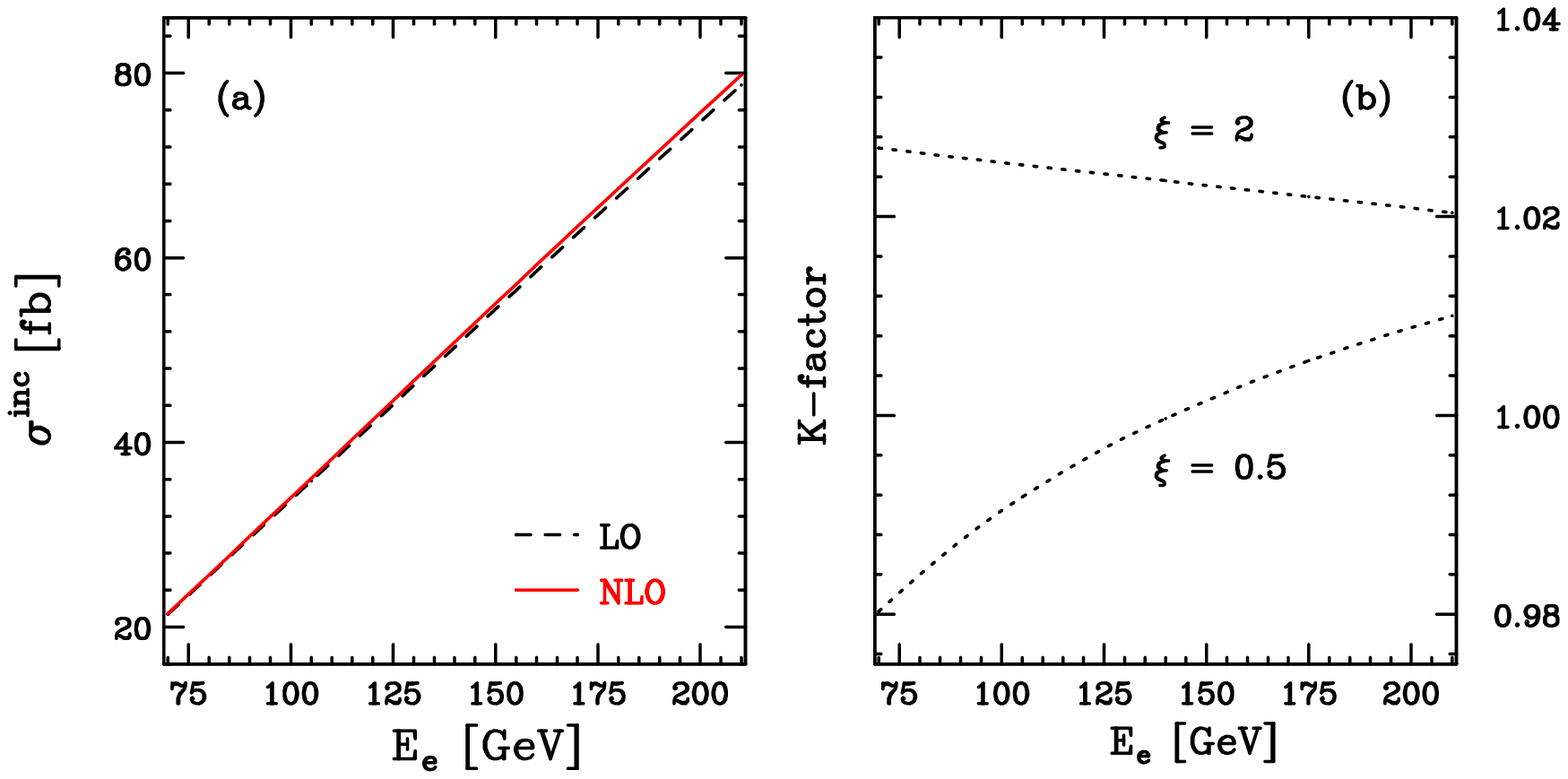}
\caption{
Panel~(a) shows the dependence of the $\ncp$ cross section, $\sigma_\mr{NC}^\mr{inc}$, at the LHeC on the energy of the electron beam at LO (black dashed) and NLO (red solid) for $M_H=120$~GeV. In panel~(b) the corresponding $K$-factor as defined in Eq.~(\ref{eq:kfac}) is shown for different values of the factorization and renormalization scales, 
$\mur=\muf=\xi\,Q$.    }
\label{fig:nc-elep-dep}
\eec
\end{figure}

The theoretical uncertainties of the $\ncp$ cross section at the LHeC are very similar to those encountered in the CC production mode. 
Figure~\ref{fig:nc-scale-dep} 
shows the scale dependence of the NC cross section,  $\sigma_\mr{NC}^\mr{inc}$, at the LHeC within the cuts of Eqs.~(\ref{eq:rjd-cut})~and~(\ref{eq:ptjet-cut}) as function of the scale parameter $\xi$, defined by Eq.~(\ref{eq:scale}). As before, we use $\mu_0=Q$ a default scale.  
In Fig.~\ref{fig:nc-mh-dep} 
and Fig.~\ref{fig:nc-elep-dep}, 
$\sigma_\mr{NC}^\mr{inc}$ is shown as a function of the Higgs boson mass and of the energy of the electron beam, respectively. The corresponding $K$-factors illustrate the associated scale uncertainties when $\mur$ and $\muf$ are varied simultaneously in the range $Q/2$ to $2Q$. Similarly to the CC case, the NLO-QCD corrections are small, but increase with $M_H$, while they decrease when $E_e$ goes up. 

In Ref.~\cite{Han:2009pe}, selection cuts have been presented for separating the NC Higgs signal from the leading background processes, in particular $e^-p\to e^-b\bar b j+X$. We apply the same set of cuts in addition to the inclusive cuts of Eqs.~(\ref{eq:rjd-cut})~and~(\ref{eq:ptjet-cut}), assuming $M_H=120$~GeV and $E_e=140$~GeV.  Any jet has to be located within the rapidity range 
\beq
|y^\mr{jet}|<5\,.
\eeq
Furthermore, the tagging jet has to exhibit a transverse momentum larger than
\beq
p_T^\mr{tag}>30~\mr{GeV}\,.
\eeq
The scattered electron is supposed to obey 
\beq
p_T^\mr{e}>30~\mr{GeV}\;\;\mr{and}\;\;
|\eta^\mr{e}|<5\,,
\eeq
while for the decay products of the Higgs boson we require
\beq
p_T^{\dec}>30~\mr{GeV}\;\;\mr{and}\;\;
|\eta^{\dec}|<2.5\,. 
\eeq
With this set of selection cuts, the NLO-QCD corrections to the $\ncp$ cross section, $\sigma_\mr{NC}^\mr{cuts}$, are very small, amounting to about 2\% for $\mur=\muf=Q$. At NLO, $\sigma_\mr{NC}^\mr{cuts}$ changes by only about 2$\permil$ when the scales are varied in the range $Q/2$ to $2 Q$. Nonetheless, the shapes of kinematic distributions can receive larger corrections, as we will see below. 

\begin{figure}[!tp]
\bec
\includegraphics[width=0.9\textwidth,clip]{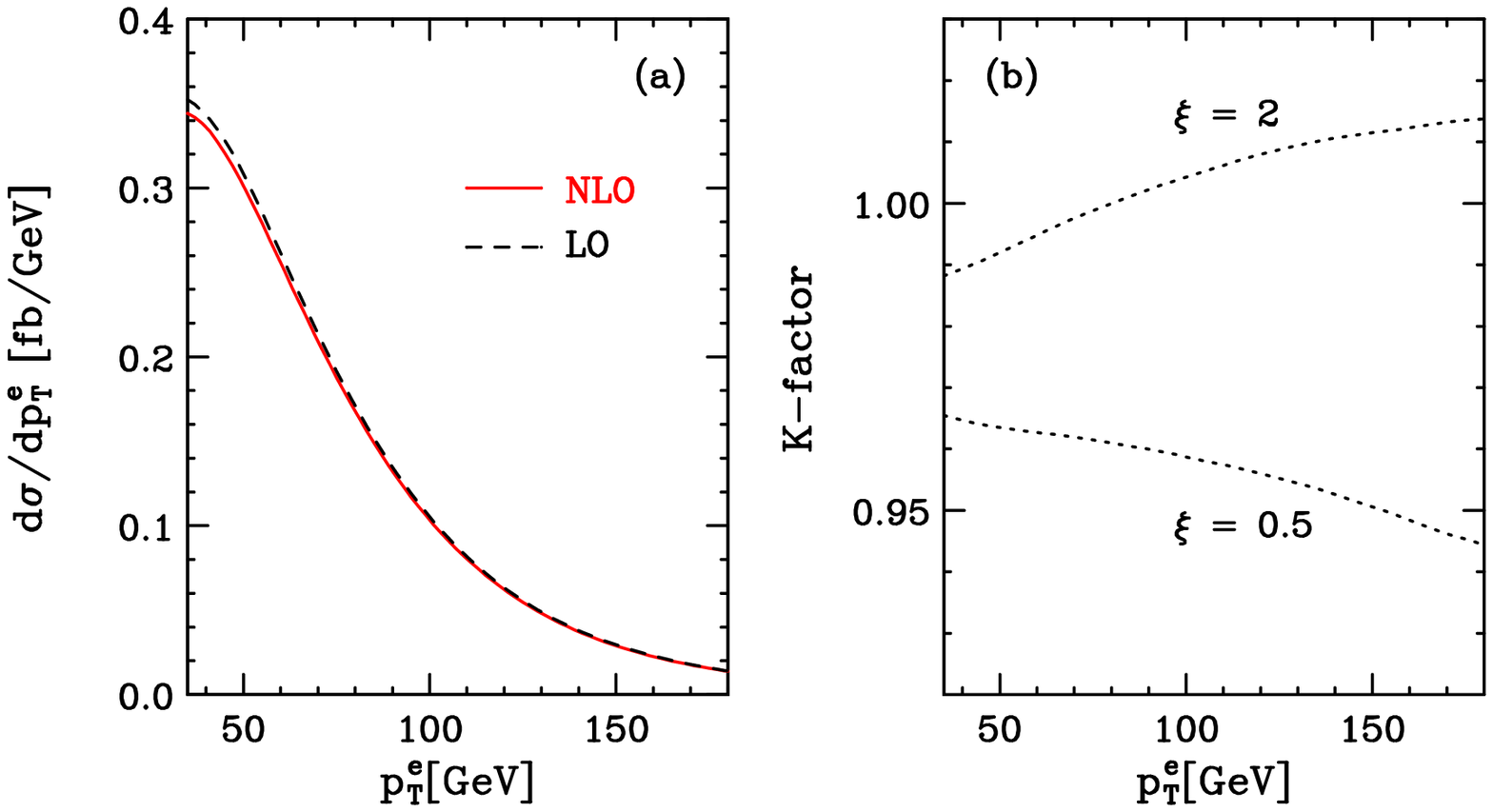}
\caption{
Transverse momentum distribution of the scattered electron in $\ncp$ at the LHeC at LO (dashed black) and NLO (solid red) [panel~(a)] and associated $K$-factor [panel~(b)] as defined in Eq.~(\ref{eq:kfac}) for different values of the factorization and renormalization scales, 
$\mur=\muf=\xi\,Q$.     }
\label{fig:nc-pt-lbr}
\eec
\end{figure}

\begin{figure}[!tp]
\bec
\includegraphics[width=0.9\textwidth,clip]{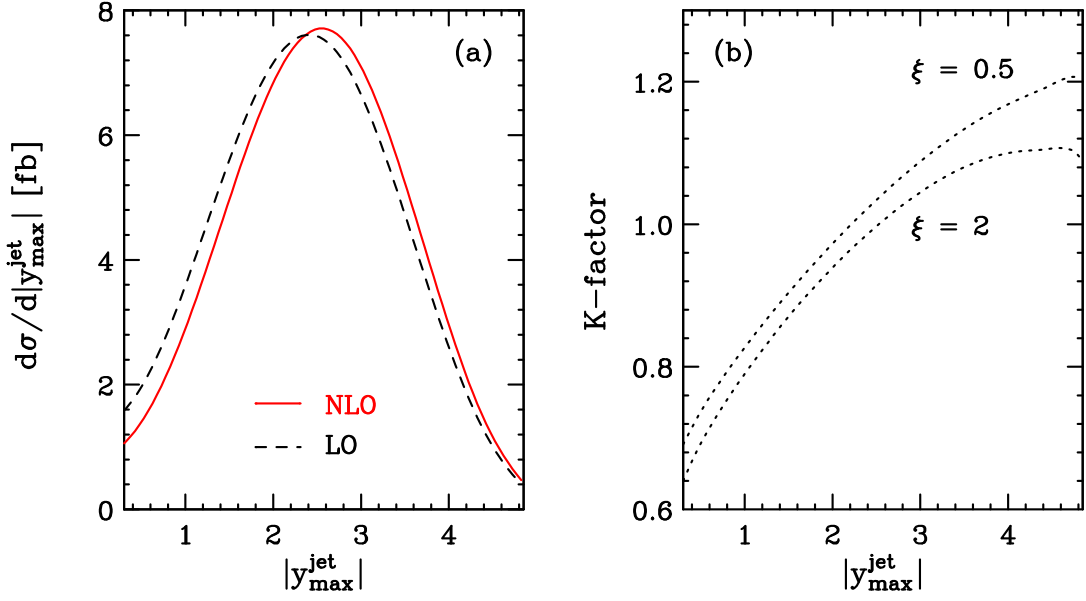}
\caption{
Rapidity distribution of the jet with the highest value of $|y^\mr{jet}|$ in $\ncp$ at the LHeC at LO (dashed black) and NLO (solid red) [panel~(a)] and associated $K$-factor   [panel~(b)] as defined in Eq.~(\ref{eq:kfac}) for different values of the factorization and renormalization scales, 
$\mur=\muf=\xi\,Q$.   }
\label{fig:nc-eta-jet}
\eec
\end{figure}

We first explore the kinematic properties of the scattered electron. Figure~\ref{fig:nc-pt-lbr}~(a)
shows the transverse-momentum distribution of the electron in the final state. The shape of this distribution is barely affected by NLO-QCD corrections. The corresponding $K$-factor is shown for $\muf=\mur=Q/2$ and $2Q$ in Fig.~\ref{fig:nc-pt-lbr}~(b), with the difference between the curves indicating the scale uncertainty of the prediction. The relative scale dependence is small, but increases with $p_T^e$. 
Similarly to $d\sigma/dp_T^e$, the rapidity distribution of the electron retains its form at NLO. Radiative corrections do have an impact on jet observables, however. In Fig.~\ref{fig:nc-eta-jet}  
the rapidity distribution of the jet with the highest value of $|y^\mr{jet}|$ is shown together with the corresponding $K$-factor. Obviously, at NLO the probability to find a jet of high rapidity is significantly larger than at LO due to the extra parton that can be emitted beyond LO. This feature is illustrated by the $K$-factor being well below one for low rapidities, but rising continually for larger values of $|y^\mr{jet}|$. The impact of scale uncertainties is quite uniform over the entire rapidity range considered, increasing slightly for large $|y^\mr{jet}|$. 

\begin{figure}[!tp]
\bec
\includegraphics[width=0.9\textwidth,clip]{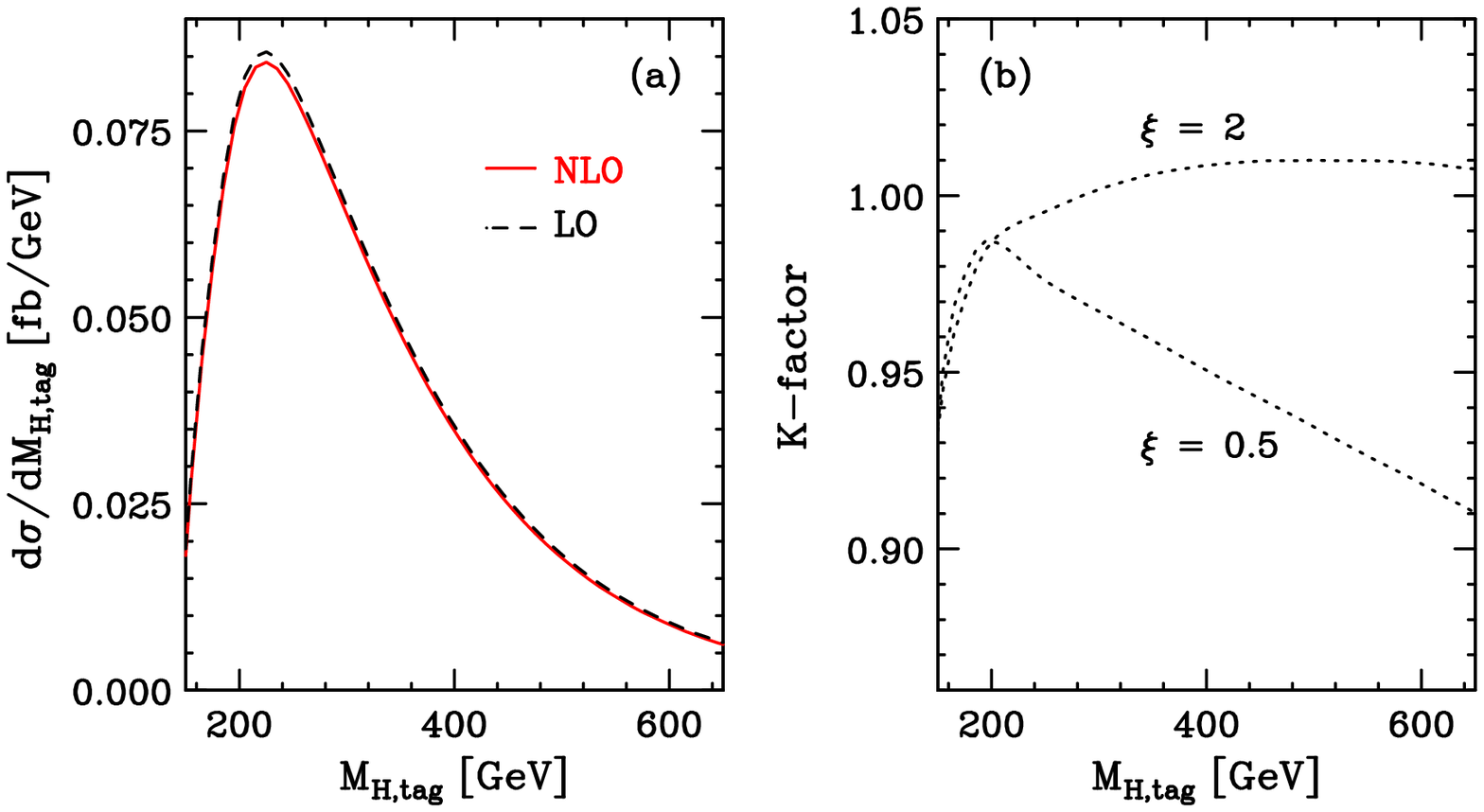}
\caption{
Invariant mass distribution of the tagging jet and the Higgs boson candidate in  $\ncp$ at the LHeC at LO (dashed black) and NLO (solid red) [panel~(a)] and associated $K$-factor   [panel~(b)] as defined in Eq.~(\ref{eq:kfac}) for different values of the factorization and renormalization scales, 
$\mur=\muf=\xi\,Q$.   }
\label{fig:nc-mhj}
\eec
\end{figure}

Figure~\ref{fig:nc-mhj}
shows the invariant mass distribution of the tagging jet and the Higgs boson candidate, $M_\mr{H,tag}$. The NLO-QCD effects on this observable are somewhat smaller than on the $|y^\mr{jet}_\mr{max}|$ distribution, but still noticeable. In  analogy to the transverse-momentum distributions discussed above, this observable is found to exhibit scale uncertainties increasing with the kinematic invariant, but still rather mild over the entire range in $M_\mr{H,tag}$ considered.

\section{Conclusions}
\label{sec:conc}
In this work, we have presented an NLO-QCD calculation of CC and NC Higgs production at an electron-proton collider. We have developed a flexible Monte Carlo program keeping full track of the kinematic properties of all particles involved in the reactions. We have studied the theoretical uncertainties being associated with the NLO-QCD predictions for inclusive cross sections as well as experimentally accessible observables within selection cuts for an optimized extraction of the Higgs signal from various backgrounds. The NLO-QCD corrections to the cross sections are small, changing the LO results by less than one percent for inclusive settings and about 3\% after optimized selection cuts have been applied.  The relative size of the QCD corrections decreases when the lepton-beam energy is increased. At NLO, the residual scale uncertainties are at the permille level, indicating that the perturbative calculation is under excellent control. These findings support the conclusion of Han and Mellado~\cite{Han:2009pe}, based on LO studies, that Higgs production in $ep$ collisions should allow for an extraction of the bottom Yukawa couplings at the LHeC with backgrounds and theoretical uncertainties being well under control.

\section*{Acknowledgments}
I am grateful to Marco Stratmann for valuable comments. This work has been supported by the Initiative and Networking Fund of the
Helmholtz Association, contract HA-101 ("Physics at the Terascale").


\begin{thebibliography}{99}

\bibitem{Duhrssen:2004cv}
M.~D\"uhrssen {\it et al.}, 
Phys.~Rev.~D~{\bf 70}, 113009 (2004) 
[hep-ph/0406323]. 
%
\bibitem{Zeppenfeld:2000td}
D.~Zeppenfeld, R.~Kinnunen, A.~Nikitenko,  and E.~Richter-Was, 
Phys.~Rev.~D~{\bf 62}, 013009 (2000) 
[hep-ph/0002036].
%
\bibitem{Rainwater:1998kj}
D.~L.~Rainwater, D.~Zeppenfeld, and K.~Hagiwara, 
Phys.~Rev.~D~{\bf 59}, 014037 (1999) 
[hep-ph/9808468].

\bibitem{Mangano:2002wn}
M.~L.~Mangano {\it et al.}, 
Phys.~Lett.~B~{\bf 556}, 50 (2003) 
[hep-ph/0210261].

\bibitem{Gabrielli:2007wf}
E.~Gabrielli {\it et al.}, 
Nucl.~Phys.~B~{\bf 781}, 64 (2007) 
[hep-ph/0702119]. 
%
\bibitem{Rainwater:2000fm} 
D.~L.~Rainwater, 
Phys.~Lett.~B~{\bf 503}, 320 (2001) 
[hep-ph/0004119].

\bibitem{Butterworth:2008iy} 
J.~M.~Butterworth, A.~R.~Davison, M.~Rubin, and G.~P.~Salam,
Phys.~Rev.~Lett.~{\bf 100}, 242001 (2008)
[0802.2470].
%
\bibitem{atlas-2009-088}
ATLAS note, ATL-PHYS-PUB-2009-088.
%
\bibitem{Plehn:2009rk}
T.~Plehn, G.~P.~Salam, and M.~Spannowsky, 
0910.5472.

\bibitem{Dainton:2006wd}
J.~B.~Dainton {\it et al.},
JINST {\bf 1}, P10001 (2006) 
[hep-ex/0603016].

\bibitem{UtaKlein09}
U.~Klein,
Talk given at the 2$^{nd}$ CERN-ECFA-NuPECC Workshop on the LHeC, Divonnes, France, September 2009. 
%
\bibitem{Kuze09}
M.~Kuze, 
Talk given at the 2$^{nd}$ CERN-ECFA-NuPECC Workshop on the LHeC, Divonnes, France, September 2009.

\bibitem{Han:2009pe}
T.~Han and B.~Mellado,
0909.2460.

\bibitem{Hagiwara:1985yu}
K.~Hagiwara and D.~Zeppenfeld,
Nucl.\ Phys.\ {\bf B274}, 1 (1986).
%
\bibitem{Hagiwara:1988pp}
K.~Hagiwara and D.~Zeppenfeld,
Nucl.\ Phys.\  {\bf B313}, 560 (1989).

\bibitem{Figy:2003nv}
T.~Figy, C.~Oleari, and D.~Zeppenfeld,
Phys.\ Rev.\ D {\bf 68}, 073005 (2003)
[hep-ph/0306109]. 
%
\bibitem{Berger:2004pca}
E.\ L.\ Berger and J.\ Campbell,
Phys.\ Rev.\ D {\bf 70}, 073011 (2004) 
[hep-ph/0403194].
%
\bibitem{Ciccolini:2007ec}
M.~Ciccolini, A.~Denner, and S.~Dittmaier,
Phys.\ Rev.\  D {\bf 77}, 013002 (2008)
[0710.4749].


\bibitem{Catani:1996vz}
S.~Catani and M.~H.~Seymour,
Nucl.\ Phys.\  {\bf B485}, 291 (1997)
[Erratum-ibid.\  {\bf B510}, 503 (1997)]
[hep-ph/9605323].

\bibitem{Stelzer:1994ta}
T.~Stelzer and W.~F.~Long,
Comput.\ Phys.\ Commun.\  {\bf 81}, 357 (1994)
[hep-ph/9401258]. 

\bibitem{Maltoni:2002qb}
F.~Maltoni and T.~Stelzer,
JHEP {\bf 0302}, 027 (2003)
[hep-ph/0208156]. 
%
\bibitem{Alwall:2007st}
J.\ Alwall {\it et al.}, 
JHEP {\bf 0709}, 028 (2007)
[0706.2334].

\bibitem{Martin:2009iq}
A.~D.~Martin, W.~J.~Stirling, R.~S~Thorne, and G.~Watt, 
Eur.~Phys.~J.~C~{\bf 63}, 189 (2009), 
[0901.0002]. 

\bibitem{Catani:1992zp}
S.~Catani, Yu.~L.~Dokshitzer, and B.~R.~Webber,
Phys.\ Lett.\ B  {\bf 285}, 291 (1992).  
%
\bibitem{Catani:1993hr}
S.~Catani, Yu.~L. Dokshitzer, M.~H.~Seymour, and B.~R.~Webber,
Nucl.\ Phys.\  {\bf B406}, 187 (1993).
%
\bibitem{Ellis:1993tq}
S.~D.~Ellis and D.~E.~Soper, Phys.\ Rev.\ D
{\bf 48}, 3160 (1993)
[hep-ph/9305266].
%
\bibitem{Blazey:2000qt}
G.~C.~Blazey {\it et al.},
hep-ex/0005012.

\bibitem{Jager:2006zc} 
B.~J\"ager, C.~Oleari, and D.~Zeppenfeld,
JHEP {\bf 0607}, 015 (2006) 
[hep-ph/0603177].
%
\bibitem{Jager:2006cp}
B.~J\"ager, C.~Oleari, and D.~Zeppenfeld,
Phys.\ Rev.\ D {\bf 73}, 113006 (2006)
[hep-ph/0604200].
%
\bibitem{Bozzi:2007ur} 
G.~Bozzi, B.~J\"ager, C.~Oleari, and D.~Zeppenfeld,
Phys.\ Rev.\ D {\bf 75}, 073004 (2007)
[hep-ph/0701105].
%
\bibitem{Jager:2009xx}
B.~J\"ager, C.~Oleari, and D.~Zeppenfeld, 
Phys.~Rev.~D~{\bf 80}, 034022 (2009) 
[0907.0580].

\bibitem{Pumplin:2002vw} 
J.~Pumplin {\it et al.}, 
JHEP {\bf 0207}, 012 (2002)
[hep-ph/0201195].


\end{thebibliography}
\end{document}